\documentclass[12pt]{article}
\usepackage{epsfig}
\usepackage{psfrag}
\usepackage{latexsym}
\usepackage{amssymb}
\usepackage{amsmath}
\usepackage{mathrsfs}

\def\be{\begin{equation}}
\def\ee{\end{equation}}
\def\bc{\begin{center}}
\def\ec{\end{center}}
\def\bea{\begin{eqnarray}}
\def\eea{\end{eqnarray}}
\newcommand{\ba}{\begin{array}{c}}
\newcommand{\bad}{\begin{array}{ccc}}
\newcommand{\ea}{\end{array}}
\def\nn{\nonumber}
\def\dd{\displaystyle}

\begin{document}
\begin{titlepage}
\vspace*{-1cm}
\phantom{hep-ph/***} 

\hfill{DFPD-09/TH/11}
%\hfill{CERN-PH-TH/2006-***}

\vskip 2.5cm
\begin{center}
{\Large\bf Tri-bimaximal Neutrino Mixing from $A_4$ and $\theta_{13} \sim \theta_C$}
\end{center}
\vskip 0.2  cm
\vskip 0.5  cm
\begin{center}

{\large Yin Lin}~\footnote{e-mail address: yinlin@pd.infn.it}
\\
\vskip .1cm
Dipartimento di Fisica `G.~Galilei', Universit\`a di Padova 
\\ 
INFN, Sezione di Padova, Via Marzolo~8, I-35131 Padua, Italy
\\
\end{center}
\vskip 0.7cm
\begin{abstract}
\noindent
It is a common believe that, if the Tri-bimaximal mixing (TBM) pattern is explained by
vacuum alignment in an $A_4$ model, only a very small reactor angle, say $\theta_{13} \sim
O(\lambda^2_C)$ being $\lambda_C\equiv \theta_C$ the Cabibbo angle, can be accommodated.
This statement is based on the assumption that all the flavon fields acquire VEVs at
a very similar scale and the departures from exact TBM arise at the same perturbation level.
From the experimental point of view, however, a relatively large value $\theta_{13} \sim O(\lambda_C)$
is not yet excluded by present data. In this paper, 
we propose a Seesaw $A_4$ model in which the previous assumption
can naturally be evaded. The aim is to describe a $\theta_{13} \sim O(\lambda_C)$
without conflicting with the TBM prediction for $\theta_{12}$ which
is rather close to the observed value (at $\lambda^2_C$ level). 
In our model the deviation of the atmospherical angle from maximal is subject to the sum-rule:
$\sin ^2 \theta_{23} \approx 1/2 + \sqrt{2}/2 \cos \delta \, \sin \theta_{13}$
which is a next-to-leading order prediction of our model.

\end{abstract}
\end{titlepage}
\setcounter{footnote}{0}
\vskip2truecm

\section{Introduction}
The present data \cite{data}, at 1$\sigma$, on solar and atmospherical angles:
\be
\theta_{12}=(34.5\pm 1.4)^o~~~,~~~~~~~\theta_{23}=(42.3^{+4.4}_{-3.5})^o~~~,
\label{angles}
\ee
are fully compatible with the TBM matrix:
\be
U_{\text{TB}}=\left(
\begin{array}{ccc}
\sqrt{2/3}& 1/\sqrt{3}& 0\\
-1/\sqrt{6}& 1/\sqrt{3}& -1/\sqrt{2}\\
-1/\sqrt{6}& 1/\sqrt{3}& +1/\sqrt{2}
\end{array}
\right)~,
\label{TB}
\ee
which corresponds to $\sin^2\theta_{12}=1/3~(\theta_{12}=35.3^o)$ and $\sin^2\theta_{23}=1/2$.
The TBM pattern also predicts $\theta_{13}=0$, however,
recent analysis based on global fits of the available data leads to hints for $\theta_{13} >0$
\cite{FogliIndication, MaltoniIndication}. 
Differently from solar or atmospherical mixing angles, the reactor one is less constrained and
its value can still be relatively large, even at 1$\sigma$ level, say $\sim \lambda_C$:
$$ \sin^2{\theta_{13}}=0.016\pm0.010 ~~~ \cite{FogliIndication}~, \qquad
\sin^2{\theta_{13}}=0.010^{+0.016}_{-0.011} ~~~ \cite{MaltoniIndication}~~~.$$
For this reason the future experimental sensitivity on the
reactor angle is fundamental for theoretical understanding of the TBM, 
which, due to its highly symmetric structure,
strongly suggests an underlying non-abelian flavour symmetry.
A natural and economical class of models based on $A_4$ flavour symmetry \cite{TB1, TB2, 
A4Lin1, A4Lin2, A4AltarelliMeloni}  has been proposed in describing TBM pattern. 
There are also more involved models based on other discrete groups
\cite{TBTprime, TBdiscrete}  or continuous flavour symmetries \cite{TBcontinuous}. 
However, if the TBM pattern results from a spontaneously broken flavour
symmetry, higher order corrections should introduce deviation from exact TBM, 
generally all of the same order.
Since the experimental departures of $\theta_{12}$ from its Tri-bimaximal value are at most of order $\lambda_C^2$, a future observed value of $\theta_{13}$ near its present upper bound should 
impose severe constraints on model buildings \cite{nonzeroUe3}. 
If this is the case, one apparently has to renounce the
nice symmetric nature of $\theta_{12}$ and simply imagines that its Tri-bimaximal value
may be completely accidental \cite{S4+theta13}. 

In this paper we will show that the TBM pattern 
can be explained by $A_4$ symmetry without necessarily implying a very 
small value for $\theta_{13}$. The fundamental new ingredient of our construction
is to allow a moderate hierarchy, of order $\lambda_C$, between the VEVs of flavon fields 
of the charged lepton and the neutrino 
sectors. The paper is organized as follows. In the next section, 
we characterize some general conditions under which 
a hierarchy between VEVS of flavon fields of different sectors can be accommodated without fine tuning.
In the section 3, we introduce the field content of our Seesaw model based on $A_4 \times Z_3 \times Z_4$ flavour group with great emphasis on vacuum alignment and its stability. 
In section 4, we explain how
the charged lepton hierarchy can be reproduced by a particular symmetry breaking 
pattern of $A_4$. In section 5, we describe the neutrino mass at leading order 
by Seesaw mechanism and obtain an exact TBM at this level.
Then in section 6 we include all subleading corrections up to 
terms suppressed by $1/\Lambda^2$ to our model and analyze possible
deviation  from TBM. In the end, in section 7, we comment on other possible
phenomenological consequences
of our model and conclude.

\section{General Consideration}
The difficulty in the standard formulation of $A_4$ models \cite{TB1, TB2} 
in generating a relatively large value of 
$\theta_{13}$ is related to the vacuum alignment
problem which plays a fundamental rule in order to naturally describe
the TBM pattern from spontaneously broken flavour symmetries.
The group $A_4$ (see Appendix A) has two important subgroups: $G_S$, which is a reflection subgroup
generated by $S$ and $G_T$, which is the group generated by $T$, isomorphic to $Z_3$.
$A_4$ can be spontaneously broken by VEVs of two sets of flavon fields, $\Phi$ for
the neutrino sector and $\Phi'$ for the charged lepton sector.
The direction of $\langle \Phi \rangle$ should leave the subgroup $G_S$ unbroken leading to the TBM. However one generally has two options for the the alignment of $\Phi'$. 
$\langle \Phi' \rangle$ can be such that $G_T$ is preserved leading
to diagonal charged lepton masses but their hierarchy is usually generated 
by an independent Froggatt-Nielsen (FN) mechanism \cite{FN}.
The second option is to consider a vacuum alignment of $\Phi'$
which entirely breaks $A_4$ and in this case the mass hierarchy is directly related to $\langle \Phi'
\rangle /\Lambda$, being $\Lambda$ the cut-off scale, 
without an extra FN component \cite{A4Lin1, A4Lin2, A4AltarelliMeloni, S3}. 
A natural mechanism for the vacuum alignment of $\Phi$ and $\Phi'$
in different directions requires the existence of an Abelian factor $G_A$ in addition to $A_4$. 
The aim of $G_A$ is to guarantee the following decomposition of the scalar potential as: 
\be
\label{general-V}
V(\Phi, \Phi') = V_\nu(\Phi)+V_e(\Phi')+V^{\rm{NLO}}(\Phi, \Phi')+\cdots ~
\ee
where we see that the interaction term between $\Phi$ and $\Phi'$ appears
from next-to-leading order (NLO). 
We will refer this situation as a ``partial'' separation in the scalar potential 
which is tightly related on the fact that only one of the sets $\Phi$ and $\Phi'$ is charged
under $G_A$, a standard choice in the literature \cite{TB1, TB2, A4AltarelliMeloni}.
At leading order, the two scalar sectors are actually separated, however, 
the vacuum alignments are affected by NLO corrections 
encoded in $V^{\rm{NLO}}(\Phi, \Phi')$.
The order of magnitude of the corrections to the VEVs $\langle \Phi \rangle$ and $\langle \Phi' \rangle$ depends on $\langle \Phi \rangle /\Lambda$ 
and $\langle \Phi' \rangle /\Lambda$ and they are subject to some conditions.
First of all, the corrections to the Tri-bimaximal value of $\theta_{12}$ are at most of order 
$\lambda_C^2$.  Furthermore, the corrections to $\langle \Phi' \rangle$ are required to be
smaller than $m_\mu/m_\tau \sim O(\lambda_C^2)$ or more restrictively smaller 
than $m_e/m_\mu \sim O(\lambda_C^3)$; otherwise, the generated charged lepton hierarchy
should not be stable. These conditions shall translate to upper bounds on the scale of 
flavour symmetry breaking with respect of the cut-off scale:
\begin{equation}
\langle \Phi \rangle /\Lambda , \langle \Phi' \rangle /\Lambda \lesssim \lambda_C^2
\label{cond1}	
\ee 
$\langle \Phi' \rangle /\Lambda$ $\lesssim \lambda_C^2$. In conclusion, a value of $\theta_{13}$ 
near its present experimental bound can not be described if the scalar potential 
is ``partially'' separated as quoted in (\ref{general-V}).

In this paper we will exploit the possibility of
a ``fully'' separated scalar potential which corresponds to (\ref{general-V}) with
$V^{\rm{NLO}}(\Phi, \Phi')=V^{\rm{NLO}}(\Phi)$ or $V^{\rm{NLO}}(\Phi, \Phi')=V^{\rm{NLO}}(\Phi')$. 
The ``fully'' separated scalar potential can be obtained if $G_A$ is a direct product of
two Abelian factors $G^\nu_A$ and $G^e_A$ which separately acts on $\Phi$ and $\Phi'$.
In this case, since $V_\nu(\Phi)$ and $V_e(\Phi')$ can be minimized in a completely independent way, 
even including NLO corrections, we are not necessarily subject to the strict condition (\ref{cond1}). 
In fact, it is possible to construct a completely natural model for TBM based on the
$A_4$ symmetry in which $\langle \Phi' \rangle /\Lambda \sim O(\lambda^2_C)$ 
and $\langle \Phi \rangle /\Lambda \sim O(\lambda_C)$
can be compatible with all experimental constraints. 
The model belongs the constrained $A_4$ models considered in \cite{A4Lin1, A4Lin2}
in which the leading order neutrino TBM and the charged lepton mass hierarchy are simultaneously reproduced by the vacuum alignment.
Our choice for $G_A$ in order to guarantee a ``fully'' separated scalar potential is
given by $Z_3 \times Z_4$. We are particularly interested in analyzing 
the possibility to have a relatively large value of $\theta_{13}$ without fine-tuning. 
We will show indeed that $\theta_{13}$ can be of order $\lambda_C$ 
while $\theta_{12}$ is corrected by subleading effects arising at order  $\lambda_C^2$. 
Furthermore,  deviations from TBM can be more intriguing since they obey a definite sum-rule
which can be in principle tested. 

\section{Field Content and Vacuum Alignment}

In this section we introduce the field content of the model and analyze the most general
scalar potential which is invariant under the flavour symmetry $A_4 \times Z_3 \times Z_4$.
The lepton SU(2) doublets $l_i$ $(i=e,\mu,\tau)$ are assigned to the triplet $A_4$ representation,
while the lepton singlets $e^c$, $\mu^c$ and $\tau^c$ are all invariant under $A_4$.
The neutrino sector is described by Seesaw mechanism with
3 heavy right-handed neutrinos $\nu^c_i$ which also form an $A_4$ triplet.
The symmetry breaking sector consists of the scalar fields neutral under the SM gauge group, 
divided in two sets as advanced before: $\Phi=\{\varphi_S,\xi,\zeta\}$ and $\Phi'=\{\varphi_T,\xi'\}$.
As anticipated before, in addition to $A_4$, we also have an Abelian symmetry $G_A=Z_3 \times Z_4$
which is a distinguishing feature of our construction.
All the fields of the model, together with their
transformation properties under the flavour group, are listed in Table~\ref{transform}.
We observe that $\Phi$ is charged under $Z_3$ while $\Phi'$ is charged under $Z_3'$.
\begin{table}
\centering
\begin{tabular}{|c||c|c|c|c|c||c|c|c|c|c|c|c|}
\hline
{\tt Field}& l & $e^c$ & $\mu^c$ & $\tau^c$ & $\nu^c$ & $h_u$ & $h_d$& 
$\varphi_T$ &  $\xi'$ & $\varphi_S$ & $\xi, \tilde{\xi}$ & $\zeta$ \\
\hline
$A_4$ & $3$ & $1$ & $1$ & $1$ & $3$ & $1$ &$1$ &$3$ & $1'$ & $3$ & $1$ & $1$ \\
\hline
$Z_3$ & $1$ & $1$ & $1$ & $1$ &  $\omega$ & $1$ & $1$&
$1$ & $1$ & $\omega$ & $\omega$ & $\omega^2$ \\
\hline
$Z_4$ &$1$ & $-1$ & $-i$ & $1$ & $1$ & $1$ & $-i$&
$i$ & $i$ & $1$ & $1$ & $1$ \\
\hline
\end{tabular}
\caption{The transformation properties of leptons, electroweak Higgs doublets and flavons under $A_4 \times
Z_3 \times Z_4$.}
\label{transform}
\end{table}

The vacuum alignment problem of the model can be solved by the supersymmetric driving
field method introduced in \cite{TB2}. 
This approach exploits the continuous $U(1)_R$ symmetry in the superpotential $w$
under which matter fields have $R=+1$, while Higgses and flavons have $R=0$. 
The spontaneous breaking of $A_4$ can be employed by adding to fields already
present in Table~\ref{transform} a new set of multiplets, called driving fields, with $R=2$.
We introduce a driving field $\xi_0$, fully invariant under $A_4$, and two driving fields
$\varphi_0^T$ and $\varphi_0^S$, triplet of $A_4$. The driving fields
$\xi_0$ and  $\varphi_0^S$, which are responsible  for the alignment of $\varphi_S$, 
have a charge $\omega$ under $Z_3$ and are invariant under $Z_4$.
$\varphi_0^T$ has a charged $ -1 $ under $Z_4$, invariant under $Z_3$, 
and drives a non-trivial VEV of $\varphi_T$.
The most general driving superpotential $w_d$ invariant under $A_4 \times G_A$ with $R=2$ is
a sum of two independent parts $w_d= w^\nu_d(\xi_0, \varphi_0^S, \Phi)+w^e_d(\varphi_0^T, \Phi'$) where 
\bea
w^\nu_d&=&g_1 \varphi_0^S \varphi^2_S+
g_2 \tilde{\xi} (\varphi_0^S \varphi_S)+
g_3 \xi_0 (\varphi_S\varphi_S)+
g_4 \xi_0 \xi^2+
g_5 \xi_0 \xi \tilde{\xi}+
g_6 \xi_0 \tilde{\xi}^2+g_7 M_\zeta \xi_0 \zeta  \label{wd1}\\
w^e_d&=& 
h_1 \xi' (\varphi_0^T \varphi_T)''+
h_2 (\varphi_0^T \varphi_T\varphi_T)~.\label{wd2}
\eea
The ``fully'' separated superpotential is guaranteed by $G_A=Z_3 \times Z_4$.
(\ref{wd1}) and (\ref{wd2}) gives two decoupled sets of F-terms 
for driving fields which characterize the supersymmetric minimum.
In other words, $w^\nu_d$ and $w^e_d$ independently determine the vacuum alignment of 
$\Phi$ and $\Phi'$, respectively. From (\ref{wd1}) we have:
\bea
\frac{\partial w} {\partial \varphi^S_{01}}&=&g_2\tilde{\xi} {\varphi_S}_1+
2g_1({\varphi_S}_1^2-{\varphi_S}_2{\varphi_S}_3)=0\nn\\
\frac {\partial w} {\partial \varphi^S_{02}}&=&g_2\tilde{\xi} {\varphi_S}_3+
2g_1({\varphi_S}_2^2-{\varphi_S}_1{\varphi_S}_3)=0\nn\\
\frac{\partial w}{\partial \varphi^S_{03}}&=&g_2\tilde{\xi} {\varphi_S}_2+
2g_1({\varphi_S}_3^2-{\varphi_S}_1{\varphi_S}_2)=0\nn\\
\frac{\partial w}{\partial \xi_0}&=&
g_4 \xi^2+g_5 \xi \tilde{\xi}+g_6\tilde{\xi}^2+g_7 M_\zeta \zeta
+g_3({\varphi_S}_1^2+2{\varphi_S}_2{\varphi_S}_3)=0~.
\eea
In a finite portion of the parameter space, we find the following stable solution
\begin{eqnarray}
\langle \tilde{\xi} \rangle&=&0~,~~~\langle \xi \rangle=u~,~~~\langle \zeta \rangle=v~, \nn\\
\langle \varphi_S \rangle &=&(v_S,v_S,v_S)~,~~~v_S^2=-\frac{g_4 u^2 + g_7 M_\zeta v}{3 g_3}~,
\label{solS}
\end{eqnarray}
with $u$ and $v$ undetermined. Since $\langle\tilde{\xi}\rangle=0$ 
{\footnote {Since there is no fundamental distinction between the singlets
$\xi$ and $\tilde{\xi}$ we have defined $\tilde{\xi}$ as the combination
that couples to $(\varphi_0^S \varphi_S)$ in the superpotential $w_d$.
The introduction of an additional singlet is essential to recover a non-trivial solution.}},
we have ignored the existence of $\tilde{\xi}$ in the rest of the paper.
Setting to zero the F-terms from Eq.~(\ref{wd2}), we obtain:
\bea
\frac{\partial w}{\partial \varphi^T_{01}}&=&h_1\xi' {\varphi_T}_3+
2h_2({\varphi_T}_1^2-{\varphi_T}_2{\varphi_T}_3)=0\nn\\
\frac{\partial w}{\partial \varphi^T_{02}}&=&h_1\xi' {\varphi_T}_2+
2h_2({\varphi_T}_2^2-{\varphi_T}_1{\varphi_T}_3)=0\nn\\
\frac{\partial w}{\partial \varphi^T_{03}}&=&h_1\xi' {\varphi_T}_1+
2h_2({\varphi_T}_3^2-{\varphi_T}_1{\varphi_T}_2)=0\nn
\eea
and the stable solution to these four equations is:
\be
\langle \xi' \rangle =u' \ne 0~,~~~\langle \varphi_T \rangle =(0, v_T,0)~,~~~v_T=-\frac{h_1u'}{2h_2}~,
\label{solT}
\ee
with $u'$ undetermined. The flat directions can be removed by the interplay of radiative corrections to the scalar potential and soft SUSY breaking terms.
It is worth to observe that, thanks to $G_A$, the VEV alignments (\ref{solS}) and (\ref{solT}) are independent even at NLO. 

Since the VEVs of the scalar fields in $\Phi$ ($\Phi'$) are related each other by adimensional
constants of order one, we should expect that they have a common scale indicated by
$\langle \Phi \rangle$ ($\langle \Phi' \rangle$).
However  $\langle \Phi \rangle /\Lambda$ and $\langle \Phi' \rangle /\Lambda$ 
can be in principle different
and they are subject to phenomenological constraints. 
As we will see in the next section, $\langle \Phi' \rangle$ is responsible for charged lepton
hierarchy so we have to require 
$$ \frac{m_e}{m_\mu} \sim \lambda_C^3 \lesssim \frac{\langle \Phi' \rangle} {\Lambda} 
\lesssim  \lambda_C^2 \sim \frac{m_\mu}{m_\tau} ~.$$
The superpotential $w^e_d$ is affected by non-renormalizable terms 
(see Appendix B for the detail) from the neutrino sector $\Phi$
suppressed by $1/\Lambda^2$. Requiring that the sub-leading corrections to $\langle \Phi' \rangle$ 
are smaller than $m_\mu/m_\tau \sim O(\lambda_C^2)$, we obtain the condition
$$ \frac{\langle \Phi \rangle} {\Lambda} \lesssim \lambda_C ~.$$
The vacuum alignment with a ``fully'' separated scalar potential allows a 
hierarchy between the VEVs of the scalars in different sectors 
$\langle \Phi' \rangle \ll \langle \Phi \rangle$.

Differently from $w^e_d$~, $w^\nu_d$ receives NLO corrections which are suppressed
only by $1/\Lambda$ but don't depend on the charged lepton sector $\Phi'$: 
$$\delta w^\nu_d = \frac 1 \Lambda \left [ (\varphi^S_0 \varphi_S) \zeta^2 + \xi_0 \xi \zeta^2 \right ]~.$$
One may wonder if a large VEV of $\Phi$ with $\langle \Phi \rangle/\Lambda \sim \lambda_C$
could introduce a too large correction to the leading order vacuum alignment (\ref{solS})
destroying the stability of the TBM prediction.
Fortunately, this is not the case. Since there is no fundamental distinction 
between $\zeta^2$ and $\xi$ the NLO correction $\delta w^\nu_d$ should induce
terms which have the same form of those already present in $w^\nu_d$. 
In fact, including $\delta w^\nu_d$ in the minimization, one easily find that
the $\langle \varphi_S \rangle$ receives only a small shift in the same direction of the leading
order alignment. For this reason we will no longer consider VEV shifts of $\varphi_S$ 
in the following.

%%%%%%%%%%%%%%%%%%%%%%%%%%%%%%%%%%%%%%%%%%%%%%%%%%%%%%%%%%%%%%%%%

%%%%%%%%%%%%%%%%                 TBM from S4 and the charged lepton hierarchies                %%%%%%%%%%%%%%%%

%%%%%%%%%%%%%%%%%%%%%%%%%%%%%%%%%%%%%%%%%%%%%%%%%%%%%%%%%%%%%%%%%

\section{Charged Lepton Hierarchy}

In the present section, we illustrate how a fully broken $A_4$ symmetry can generate
the charged lepton hierarchy.
The key ingredient is the alignment $\langle \varphi_T\rangle \sim (0, 1, 0)$. 
Such a VEV breaks the permutation symmetry
 of the second and third generation of neutrinos
in a maximal way in the sense that $$ \langle \varphi_T\rangle^t S_{2-3} \langle \varphi_T\rangle=0~, $$
where $$S_{2-3}= \left(
\begin{array}{ccc}
1&0&0\\
0&0&1\\
0&1&0
\end{array}
\right).$$

The $A_4$ group is fully broken
{\footnote{Similarly as explained in \cite{A4Lin1, A4AltarelliMeloni}, 
a residual symmetry $A_4 \times Z_3$ from $A_4 \times Z_4$
survives in the charged lepton sector
guaranteeing the stability of the vacuum alignment.}} in
the charged lepton sector by $\Phi'$ with the vacuum structure quoted in (\ref{solT})
and only the tau mass is generated at leading order.
The muon and electro masses are generated respectively by
$\langle \varphi_T\rangle^2 \propto (0,0,1)$ and $\langle \varphi_T\rangle^3 \propto (1,0,0)$.
Then the correct hierarchy between the charged lepton masses
$m_e \ll m_{\mu} \ll m_{\tau}$ is reproduced if we assume 
$\lambda_C^2 \lesssim \langle \Phi' \rangle / \Lambda \lesssim \lambda_C^3 $.

Since $\Phi'$ carries a charge $i$ under $Z_4$ we have to assign different $Z_4$ 
charges for lepton singlets. 
Considering only insertions of $\Phi'$, the charged lepton masses are described by $w_e$, 
given by, up to $1/ \Lambda^3$:
\bea
w_e &=& \alpha_1 \tau^c (l \varphi_T)h_d/\Lambda \nn \\
&+& \beta_1 \mu^c \xi' (l \varphi_T)'' h_d/\Lambda^2+\beta_2 \mu^c (l \varphi_T \varphi_T) h_d/\Lambda^2 \nn \\
&+& \gamma_1 e^c (\xi')^2 (l \varphi_T)' h_d/\Lambda^3+
\gamma_2 e^c \xi' (l \varphi_T \varphi_T)'' h_d/\Lambda^3+
\gamma_3 e^c (l \varphi_T \varphi_T \varphi_T)h_d/\Lambda^3 \nn
\label{wlplus}
\eea
After electroweak symmetry breaking, $\langle h_{u,d}\rangle=v_{u,d}$~, 
given the specific orientation of 
$\langle \varphi_T \rangle \propto (0, 1, 0)$~, $w_e$ 
give rise to diagonal and hierarchical mass terms for charged leptons.
Defining the expansion parameter $v_T/\Lambda \equiv \lambda^2 \ll 1$ (it is not restrictive to
consider $v_T$ to be positive)
and the Yukawa couplings $y_l$ ($l=e, \mu, \tau$) as 
\bea
y_\tau&=&|\alpha_1| \nn~, \\ 
y_\mu&=& |\beta_1u'/v_T +2 \beta_2 | \lambda^2\nn ~, \\
y_e &=& |\gamma_1(u'/v_T)^2-\gamma_2 u'/v_T-2\gamma_3 \nn~, 
\eea
the charged lepton masses are given by
\be
m_l=y_l \lambda^2 v_d~~~~~~~(l=e,\mu,\tau)~~~.
\ee

As already pointed out in the previous section and analyzed in detail in Appendix B, 
the vacuum alignment for $\varphi_T$
receives correction of order $\langle \Phi \rangle^2 / \Lambda^2 \sim \lambda^2_C$
different for each component:
$$\varphi_T =(\delta_{T1}, v_T+\delta_{T2}, \delta_{T3} )~.$$
Including correction to the vacuum alignment for $\varphi_T$, the diagonal form of 
the charged lepton mass should slightly change and small off-diagonal entries appear:
\be
m_e= \left(
\begin{array}{ccc}
m_e&m_e O(\lambda^2_C)&m_e O(\lambda^2_C)\\
m_\mu O(\lambda^2_C)&m_\mu&m_\mu O(\lambda^2_C)\\
m_\tau O(\lambda^2_C)&m_\tau O(\lambda^2_C)&m_\tau
\end{array}
\right).
\label{menew}
\ee
The transformation needed to diagonalize $m_e$ is 
$V^T_e m_e U_e = {\rm diag} (m_e, m_\mu, m_\tau)$ and the unitary matrix $U_e$ is given by
\be
U_e= \left(
\begin{array}{ccc}
1&O(\lambda^2_C)&O(\lambda^2_C)\\
 O(\lambda^2_C)&1& O(\lambda^2_C)\\
O(\lambda^2_C)& O(\lambda^2_C)&1
\end{array}
\right).
\label{Ue}
\ee 

Another source of off-diagonal correction to charged leptons comes from 
the interaction with the neutrino sector. 
In fact, the products $\xi \zeta$ and $\varphi_S \zeta$
are invariant combination under $G_A$ and we can include them 
on top of each term in $w_e$. 
However, we find that the introduction of these additional terms 
changes the charged lepton mass $m_e$ exactly in the same way as 
the corrections induced by VEV shifts  of $\varphi_T$, i.e. (\ref{menew}).
Then (\ref{Ue}) is the most general structure of the charged lepton contribution
to TB mixing.

\section{A Seesaw realization of the constrained $A_4$ model}

The masses of light neutrinos of our model is described by Seesaw superpotential
with 3 heavy right-handed neutrinos $\nu^c_i$, triplet of $A_4$.
Terms in the superpotential which contain $\nu^c$ invariant under
the flavour group are given by:
\be
w_\nu = y (\nu^c l) \zeta h^u/\Lambda + x_a\xi (\nu^c \nu^c)+x_b (\varphi_S \nu^c \nu^c) + h.c. +\cdots
\label{Seesaw}
\ee
In the heavy neutrino sector $A_4 \times Z_3 $ is broken by $ \langle \varphi_S\rangle = (v_S,v_S,v_S)$ and $\langle \xi \rangle = u$ down to $G_S$ 
(with $Z_4$ unbroken) with an accidental extra $G_{2-3}$ symmetry.
Then the residual symmetry of the right-handed neutrino masses is $G_{TB}=G_S \times G_{2-3}$.
$G_{TB}$ can be transfered to the light neutrino sector if the Dirac neutrino mass
commute its generators. This is in fact the case.
After electroweak and $A_4$ symmetry breaking from (\ref{Seesaw}) we obtain the following 
leading contribution to the Dirac and Majorana masses:
\be
m^D_0=\left(
\begin{array}{ccc}
1& 0& 0\\
0& 0& 1\\
0& 1& 0
\end{array}
\right)yv_u \frac v \Lambda~, \qquad
M=\left(
\begin{array}{ccc}
a+2 b& -b& -b\\
-b& 2b& a-b\\
-b& a-b& 2 b
\end{array}
\right)u~,
\label{mnu0}
\ee
where
\be
a\equiv x_a ~~~,~~~~~~~b\equiv x_b\frac{v_S}u~~~.
\label{ad}
\ee
We immediately see that $[m^D_0,S]=0$.
The leading order lepton mixing matrix is entirely encoded in the right-handed neutrino mass matrix $M$
which is diagonalized by the transformation:
\be
U_0^\dagger M U^*_0 ={\tt diag}(|a+3b|,|a|,|a-3b|)u~,
\ee
with $U_0=U_{\text{TB}} \Omega$ where  $\Omega=\rm{diag} \{ e^{i \phi_1/2},e^{i \phi_2/2},
 i e^{i \phi_3/2} \}$ and
$\phi_1,\phi_2,\phi_3$ are respectively phases of $a+3b$, $a$, $a-3b$.
Naturally $\phi_1$ and $\phi_3$ depend on $\phi_2$ and $\Delta$, 
the relative phase between $a$ and $b$.
The light neutrino masses are given by the type I Seesaw mechanism: $m_\nu= (m_0^D)^T M^{-1} m^D_0$ which is invariant under $G_{TB}$ and then also diagonalized by $U_0$
{\footnote{The overall phase appearing in the Dirac neutrino mass $m^D_0$ can be absorbed 
by the redefinition of $\phi_2$ and there are only two independent Majorana phases.}}.
Denoting the physical masses of $\nu^c_i$ as $M_1=|a+3b|$, $M_2=|a|$ and $M_3=|a-3b|$,
we obtain $$U_0^T m^0_\nu U_0= \left|\frac{y v_u v}{\Lambda}\right|^2 {\tt diag} \{ \frac 1{M_1},
\frac 1{M_2},\frac 1{M_3} \}={\tt diag}\{ m_1, m_2, m_3 \}~~~.$$
$m_2 > m_1$ implies $t \equiv |3b|/|a| > -2 \cos \Delta$ and
in principle both normal and inverted hierarchies in the neutrino spectrum
can be reproduced. The normal hierarchy is realized for  $t/2 \le \cos \Delta \le 1$
whereas an inverted spectrum requires $-t/2 < \cos \Delta \le 0$.
The ratio $r = \Delta m^2_{\rm{sun}}/ \Delta m^2_{\rm{atm}}$
(where $\Delta m^2_{\rm{sun}} = m^2_2-m^2_1$ and 
$\Delta m^2_{\rm{atm}}=|m_3^2-m_1^2|$) is given in our model by:
\be
r=\frac{(t+2 \cos \Delta)(1+t^2-2t \cos \Delta)}{4\cos \Delta}~.
\ee
One can show that for the normal hierarchy, a small value of $r \approx 1/30$ can be
reproduced only for $\cos \Delta \approx t \approx 1$.
In particular, a normal ordered spectrum can never be degenerate.
Then we can expand $t =1+\delta t$ with $\delta t \ll t$ obtaining the following 
approximate spectrum:
\be
m_1 \approx \sqrt{\Delta m^2_{\rm{sum}}/3}~,~~~
m_2 \approx 2 m_1~,~~~
m_3 \approx \sqrt{  \Delta m^2_{\rm{atm}} - \Delta m^2_{\rm{sum}}/3}~.
\ee
The inverted hierarchy can be realized only for $t \approx - 2 \cos \Delta $
and in this case we can expand $ \cos \Delta = -t/2 +\delta t'$ with $\delta t' \ll t$.
Expressing $\delta t$ in function of $r$
we obtain
\bea
m^2_1&=&\Delta m^2_{\rm{atm}} \left[ 1+\frac {1}{2t^2} +\left( \frac 1 {t^2} - \frac{1}{{1+2t^2}} \right)r \right] \nonumber \\
m_2^2 &=&\Delta m^2_{\rm{atm}} \left[ 1+\frac {1}{2t^2} +\left(1+ \frac 1 {t^2} - \frac{1}{{1+2t^2}} \right)r \right] \nonumber \\
m^2_3&=&\Delta m^2_{\rm{atm}} \left[ \frac {1}{2t^2} +\left( \frac 1 {t^2} - \frac{1}{{1+2t^2}} \right)r \right] 
\nonumber ~.
\eea

In principle, the previous expansion is valid also for a 
degenerate spectrum realized by $t \ll 1$ which is, however, parametrically fine tuned
\footnote {The fine-tuning required in order to reproduce a small $r$ becomes 
more severe if we include in $w_\nu$ also the five-dimensional operator
$lh^ulh^u/\Lambda'$ which leads to a mass matrix structure similar
to the term $\xi \nu^c \nu^c$. Indeed if the Weinberg operator 
has a cutoff scale $\Lambda' \sim \Lambda$, its contribution becomes larger than
the Seesaw one. This situation is equivalent to go to the limit $a \gg b$ and then it is disfavored.
In order to avoid this problem we will assume that the lepton number is violated only
by Majorana mass term up to $\Lambda$. In other words, we require $\Lambda' \gg \Lambda$
and a direct five-dimensional operator can be neglected.}
 in our model.

\section{Deviation from TBM and $\theta_{13} \sim \lambda_C$}

In this section we show how a relatively large reactor
angle, say $\theta_{13} \sim \theta_C$, can naturally arise in our model,
without conflicting with the precise value of $\theta_{12}$ predicted by TBM. 
The neutrino mass described in the previous section
predicts an exact TBM. Including sub-leading contributions dictated by 
higher-dimensional operators, the leading order lepton mixing matrix should be modified.
As we shall see in a moment, not all deviations from TBM arise at the same 
perturbation level, this is one of the most important feature of the model.
We find that the NLO corrections generate a non-vanishing reactor angle which is
correlated with deviation of atmospherical angle from maximal. While the corrections to 
solar angle appear only at next-to-next to leading order (NNLO).

First of all we focus on higher order corrections to the right-handed Majorana neutrino mass
up to terms suppressed by $1/\Lambda^2$. At NLO, there is only one additional contribution 
to heavy Majorana mass: $ \zeta^2 \nu^c \nu^c /\Lambda$.
Since $\zeta^2$ has exactly the same property of $\xi$, this term
can be absorbed by a redefinition of $a$.
The NNLO contributions arise from adding
the products $\xi \zeta$ and $\varphi_S \zeta$, invariant combination under $G_A$, 
on top of the leading order terms. 
In this case, not all the corrections have the same
structure of the terms already present in $w_\nu$ and consequently cannot be regarded
as small shifts of $a$ and $b$, for example 
$(\nu^c \nu^c)' (\varphi_S \varphi_S)''$ and $(\nu^c \nu^c)'' (\varphi_S \varphi_S)'$.
However these terms can be absorbed by parameters $y_1$ and $y_2$ in the 
NLO correction to the Dirac mass $\delta m^D$ as will be clear in a moment.

Now we move to consider the correction to Dirac neutrino mass: $\delta m^D$ 
beginning with terms suppressed by $1/\Lambda^2$.
There are many independent terms of the type $(\nu^c l \varphi \varphi) h^u$,
 with $\varphi \in \{ \varphi_S, \xi \}$,
invariant of $A_4$ which contribute to $\delta m^D$ at this order:
\bea
\delta w_\nu &=& h^u \frac{y_1}{\Lambda^2} (\nu^c l)' (\varphi_S \varphi_S)''+ 
 h^u \frac{y_2}{\Lambda^2} (\nu^c l)'' (\varphi_S \varphi_S)'+
h^u \frac{y_3}{\Lambda^2} \nu^c (l \varphi_S)_A \xi + \nn \\
&+& h^u \frac{y'}{\Lambda^2} (\nu^c l)_1 (\varphi_S \varphi_S)_1+ 
h^u \frac{y''}{\Lambda^2} (\nu^c l) \xi^2+
h^u \frac{y'_2}{\Lambda^2} \nu^c (l \varphi_S)_S \xi~.
\label{deltanu}
\eea
Observe that the operators with coefficients $y', y'', y'_2$ give
contribution to Dirac mass matrix in a form invariant 
under $G_{TB}$ exactly as right-handed neutrino mass.
Then these corrections can be adsorbed into a redefinition
of the leading-order coefficients.
The relevant correction to the Dirac mass comes from
the first three terms in Eq.~(\ref{deltanu}) and has the following form:
\be
\delta m^D=\left(
\begin{array}{ccc}
0 & y_1+\tilde{y}_3& y_2-\tilde{y}_3\\
y_1-\tilde{y}_3& y_2& \tilde{y}_3\\
y_2+\tilde{y}_3& -\tilde{y}_3& y_1
\end{array}
\right) v_u \frac {v_S^2} {\Lambda^2}~,
\ee
where $y_1,y_2,\tilde{y}_3 \equiv y_3 u/v_S$ are generally complex number of order $1$.
Before discussing the important consequence when we include 
the NLO correction to the Dirac neutrino mass , 
we comment possible NNLO effects on $m^D$.
Here the NNLO contributions are suppressed by $1/ \Lambda^3$
and they are of the type $(\nu^c l \zeta^2 \varphi) h^u$.
All these terms can be absorbed by a redefinition of 
$y_3$ and $y'_2$, then we can forget them in the following
analysis.

In order to find the correction to the leading neutrino mixing matrix 
$U_0=U_{\text{TB}} \Omega$, it is convenient to define
$$\hat{m}^D=U^\dagger_0 m^D U_0~,$$
where $m^D=m^D_0+\delta m^D $.
The light neutrino mass is then formally given by
$$m_\nu =U_0 \hat{m}_\nu U^T_0$$
where $\hat{m}_\nu \equiv (\hat{m}^D)^T M^{-1}_{\rm{diag}} \hat{m}^D$
with $M^{-1}_{\rm{diag}}={\tt diag} \{ 1/ M_1,  1/ M_2 ,  1M_3 \}$.
If $\hat{m}_\nu$ can be diagonalized by the unitary matrix $\delta U \sim I$ as
$$ \delta U \hat{m}_\nu \delta U^T = {\tt diag}\{ \hat{m}_1, \hat{m}_2, \hat{m}_3 \}~,$$
where $\hat{m}_i \approx m_i$ ,
the full PMNS mixing matrix will be given by 
\be
U_{\rm{PMNS}}=U^\dagger_eU_0 \delta U ~.
\label{Up}
\ee
In our case, the matrix $\hat{m}^D$ has a very simple expression:
\be
\hat{m}^D \approx \left(
\begin{array}{ccc}
1& 0& e^{i \phi_{31}} c_+ \epsilon \\
0& 1& 0\\
e^{i \phi_{31}} c_- \epsilon & 0& -1
\end{array}
\right)yv_u \frac v \Lambda~,
\ee
where $\phi_{31}= (\phi_3 -\phi_1)/2$, $c_{+(-)}=i \sqrt{3}/2 (y_2-y_1+(-)2 \tilde{y}_3)$
and $\epsilon = v_S^2/(v \Lambda) \sim \lambda_C$. Then we get
\be
\hat{m}_\nu = \left(
\begin{array}{ccc}
m_1& 0& e^{i \phi_{31}} (c_+ m_1+c_- m_3) \epsilon \\
0& m_2 & 0\\
e^{i \phi_{31}} (c_+ m_1+c_- m_3) \epsilon & 0& m_3
\end{array}
\right)+O(\epsilon^2)~.
\ee
This result means that a correction $(\delta U)_{13} \sim \lambda_C$
can be present and we can expect that a deviation of $\theta_{12}$ from it Tri-bimaximal value
arises only at order $\lambda_C^2$. However, observe that
if $m_1 \approx m_3$ i.e. the spectrum becomes degenerate, a fine-tuning 
will be required in order to reproduce a small $(\delta U)_{13}$.
From this viewpoint, a degenerate spectrum is disfavored 
if we require that the deviation from TBM is naturally small.

Forgetting for a moment $U_e$ which arises only at NNLO, 
from Eq.~(\ref{Up}), one find that
\be
U_{e 3} = \sqrt{\frac 23} e^{i \phi_{13}} (\delta U)_{13} \,,\qquad
U_{\mu 3} = -\frac 1{\sqrt{2}}+\sqrt{\frac 16} e^{i \phi_{13}} (\delta U)_{13} 
\ee
and $U_{l2}$, $l=e, \mu, \tau$, remain unchanged.
As a result, the solar angle $\theta_{12}$ remains rather close to its Tri-bimaximal value. 
However $(\delta U)_{13}$
simultaneously induces a departure of $\theta_{13}$ and of $\theta_{23}-\pi /4$ from zero.
Defining $\delta'$ as the phase of $(\delta U)_{13}$, the CP-violating
Dirac phase is given by $-\delta=\delta'+\phi_{13}$. 
Since $\sin \theta_{13}=\sqrt{2/3} |(\delta U)_{13}|$, the deviation of
the atmospherical angle from maximal is subject to the following sum-rule:
\be
\sin ^2 \theta_{23} = \frac {|U_{\mu 3}|^2}{1-|U_{e3}|^2}
\approx \frac 12 + \frac {\sqrt{2}}{2} \cos \delta\, \sin \theta_{13}+ O(\theta^2_{13}),
\ee
this is a prediction of our model. 
This is a special feature of the present Seesaw $A_4$ model. The presence 
of the abelian factor $G_A$ in our model, not only allows a relatively large value of $\theta_{13}$, at $\theta_C$ level, also strongly suppresses possible higher order contributions
giving rise correlation between them.

Independently from the Seesaw sector, TBM and in particular the solar angle 
receives corrections from charged lepton sector.
Adopting the standard parametrization of $U_{\rm{PMNS}}$, from (\ref{Up}) and (\ref{Ue})
one finds that all the mixing angles receive a correction of order $\lambda^2_C$.
Then we in particular  obtain $$\sin^2 \theta_{12}=\frac 13 + O(\lambda^2_C)~.$$
As claimed in the beginning, $\theta_{13}$ can be of order $\lambda_C$ since it arises
from corrections at NLO in the neutrino sector
while $\theta_{12}$ receives corrections only of order $\lambda_C^2$ 
which are subleading effects at NNLO. 

\section{Conclusion and discussion}

In this paper we have addressed one of the most important issues in the $A_4$ realization
of TBM i.e. if a $\theta_{13} \sim \theta_C$ can be allowed without fine tuning.
We have discussed a framework, referred as constrained $A_4$ model,
in which the vacuum alignment is realized by a fully separated scalar potential. 
The model is based on the $A_4 \times Z_3 \times Z_4$ flavour symmetry and
(Type I) Seesaw mechanism. 
In the charged lepton sector, the $A_4$ group is entirely broken by
the set of scalar field $\Phi'=\{\varphi_T, \xi' \}$. The symmetry breaking parameter
$\langle \Phi' \rangle /\Lambda  \sim \lambda^2_C$ directly
controls the charged lepton mass hierarchy without requiring a $U(1)_{\rm{FN}}$ symmetry.  
In the neutrino sector, the set of scalar fields $\Phi = \{\varphi_S, \xi, \zeta \}$ breaks the $A_4$
group to its subgroup $G_S$ guaranteeing the TBM at leading order.
The symmetry breaking parameter
$\langle \Phi \rangle /\Lambda$, however, can be chosen at order of the Cabibbo angle $\lambda_C$ 
without altering the required vacuum alignment for $\Phi'$.
Moreover, a non-vanishing $\theta_{13}$ and a deviation of $\theta_{23}$ from $\pi/4$
are simultaneously generated at order $O(\lambda_C)$ leaving $\theta_{12}$ unchanged.
Subsequently, a deviation of the solar angle from its TBM value is generated 
at order $O(\lambda^2_C)$ which just corresponds to its $1\sigma$ experimental sensitivity.

The model is called constrained $A_4$ model because, differently 
from its standard formulation widely studied in literature, the NLO corrections
are also dictated by $A_4$ symmetry itself. This is another interesting feature of 
our model. There is, indeed, a correlation between 
 the deviation of $\theta_{23}$ from maximal and the value of generated $\theta_{13}$:
$\sin ^2 \theta_{23} \approx 1/2 + \sqrt{2}/2 \cos \delta \, \sin \theta_{13} + O(\theta^2_{13})$
which can be in principle tested by future experiments. 
Concerning  the neutrino spectrum, it can be either of normal hierarchy or inverted one. 
However, a degenerated spectrum is parametrically fine tuned and is disfavored 
requiring that the deviation from TBM is naturally small. For this reason, we should also 
expect that the effect of running on mixing angles is negligible. 
Since the solar angle 
has been measured more precisely than the others, its running can be potentially important 
if the neutrino spectrum were degenerate.

The corrections beyond the leading order are important not only in describing
deviations from TBM, but also give rise other interesting phenomenology.
For example, the same breaking pattern for charged lepton sector can be easily
extended to the quark sector. In this case, the $V_{\rm{CKM}}$ arises when the
correction to the vacuum alignment $\varphi_T$ is taken into account.
 Then the resulting $V_{\rm{CKM}}$ should have the same form of the unitary matrix
diagonalizing charged leptons $U_e$ given in (\ref{Ue}). 
The inclusion of the sub leading corrections can also play an important role
in explaining the baryon asymmetry of the universe (BAU) through leptogenesis \cite{TBlepto}. 
As pointed out in \cite{A4Lin2}, the generated BAU can be indeed directly
trigged by low energy phases appearing $U_{e3}$.
Moreover, the structure of $A_4$ symmetry breaking pattern can be revealed 
by other physical effects \cite{LFV}, not directly related to neutrino properties, such as
lepton flavour violating process as well as the anomalous magnetic moments and the electric dipole moments of charged leptons. Such a possibility becomes realistic if there is new physics
at a much lower energy scale around $1-10$ TeV.
All these issues merit a further and more detailed study. 

\vskip 0.5 cm
\section*{Acknowledgements}
We thank Ferruccio Feruglio for useful suggestions and for his encouragement in our work.
We thank also Guido Altarelli, Davide Meloni and Luca Merlo for useful discussions.
We recognize that this work has been partly supported by 
the European Commission under contracts MRTN-CT-2004-503369 and MRTN-CT-2006-035505.

\section*{Appendix A: The group $A_4$}

The group $A_4$ has 12 elements and four non-equivalent irreducible representations: 
one triplet and three independent singlets $1$, $1'$ and $1''$. 
Elements of $A_4$ are generated by the two generators
$S$ and $T$ obeying the relations:
\be
S^2=(ST)^3=T^3=1~~~.
\label{$A_4$}
\ee
We will consider the following unitary representations of $T$ and $S$:
\be
\begin{array}{lll}
\text{for}~1&S=1&T=1\\
\text{for}~1'&S=1&T=e^{\dd i 4 \pi/3}\equiv\omega^2\\
\text{for}~1''&S=1&T=e^{\dd i 2\pi/3}\equiv\omega
\label{singlets}
\end{array}
\ee
and for the triplet representation
\be
T=\left(
\begin{array}{ccc}
1&0&0\\
0&\omega^2&0\\
0&0&\omega
\end{array}
\right),~~~~~~~~~~~~~~~~
S=\frac{1}{3}
\left(
\begin{array}{ccc}
-1&2&2\cr
2&-1&2\cr
2&2&-1
\end{array}
\right)~~~.
\label{ST}
\ee
The tensor product of two triplets is given by $3 \times 3 = 1+1'+1''+3_S+3_A$.
From (\ref{singlets}) and (\ref{ST}), one can easily construct all multiplication rules of $A_4$. 
In particular, for two triplets $\psi=(\psi_1,\psi_2, \psi_3)$ 
and $\varphi=(\varphi_1,\varphi_2, \varphi_3)$ one has:
\bea \label{tensorproda4}
&\psi_1\varphi_1+\psi_2\varphi_3+\psi_3\varphi_2 \sim 1 ~,\nn \\
&\psi_3\varphi_3+\psi_1\varphi_2+\psi_2\varphi_1 \sim 1' ~,\nn \\
&\psi_2\varphi_2+\psi_3\varphi_1+\psi_1\varphi_3 \sim 1'' ~,\nn
\eea
  \be
   \left( 
 \ba
 2\psi_1\varphi_1-\psi_2\varphi_3-\psi_3\varphi_2 \\
 2\psi_3\varphi_3-\psi_1\varphi_2-\psi_2\varphi_1 \\
 2\psi_2\varphi_2-\psi_1\varphi_3-\psi_3\varphi_1 \\
  \ea
  \right) \sim 3_S~, \qquad
  \left( 
 \ba
 \psi_2\varphi_3-\psi_3\varphi_2 \\
 \psi_1\varphi_2-\psi_2\varphi_1 \\
 \psi_3\varphi_1-\psi_1\varphi_3 \\
  \ea
  \right) \sim 3_A~.
  \label{tensorp}
   \ee

\section*{Appendix B: Correction to Alignment of $\varphi_T$ and $\varphi_S$}

In this appendix we will study correction to the leading order alignment 
of $\varphi_S$ and $\varphi_T$ when we include higher dimensionality
operators up to the order $1/\Lambda^2$.

In our model, the correction to the driving superpotential for $\varphi_S$,
depends only on $\Phi$ at NNLO, then the obtained 
vacuum alignment $\langle \varphi_S \rangle \propto (1,1,1)$ is always stable 
since it preserves the subgroup $G_S$ of $A_4$. However a relative large 
$\langle \Phi \rangle /\Lambda \sim \lambda_C$ may have
some effects on the leading order alignment for $\varphi_T \propto (0,1,0)$. 
The products $\xi \zeta$ and $\varphi_S \zeta$
are invariant combination under $G_A$, then we can include them 
on top of each term in $w^e_d$. With the introduction of these higher dimensionality
operators, $w^e_d$ should be modified into $w^e_d+\delta w^e_d$ where
\footnote {Here we omit the term $\zeta (\xi')^2 (\varphi_S \varphi^T_0)'$
since it induces only a small shift of $u'$ and then can be included in the redefinition
of $u'$.}

$$\delta w^e_d = \frac 1 {\Lambda^2} \left [ t_1 \zeta \xi \xi'(\varphi^T_0 \varphi_T)''+ t_2\zeta \xi (\varphi^T_0 \varphi_T \varphi_T)+t_3\zeta \xi' (\varphi^T_0 \varphi_T \varphi_S)''+t_4\zeta (\varphi^T_0 \varphi_S)' (\varphi_T \varphi_T)'' \right]~.$$
The alignment for $\varphi_T$ should be shifted (the shift in $\xi'$ is needless) and we
can look for a solution that perturbs $\langle \varphi_T \rangle$
to second order in the $1/\Lambda$ expansion:
 $$\langle \xi' \rangle = u'~, \qquad \langle \varphi_T \rangle=(\delta_{T1}, v_T+\delta_{T2}, \delta_{T3} )~.$$
The minimum conditions from $w^e_d+\delta w^e_d$ become equations in the shifts $\delta v_{Ti}$:
\bea
-4 h_2 v_T \delta v_{T3}&+&\left( t_4-t_3 \frac {4h_2}{h_1} \right) \frac {v v_S}{\Lambda^2} v^2_T =0 \nn \\
2 h_2 v_T \delta v_{T2}&+&\left( t_4+t_3 \frac {4h_2}{h_1} \right) \frac {v v_S}{\Lambda^2} v^2_T 
+\left( 2t_2-t_1 \frac {2h_2}{h_1} \right) \frac {v u}{\Lambda^2} v^2_T =0 \nn \\
 -4 h_2 v_T \delta v_{T1}&+&\left( t_4+t_3 \frac {4h_2}{h_1} \right) \frac {v v_S}{\Lambda^2} v^2_T =0\nn
\eea
These equations are linear in $\delta v_{Ti}$ and can be easily solved by:
\bea
\frac{\delta v_{T3}}{v_T}&=&\left( \frac{t_4}{4h_2}- \frac {t_3}{h_1} \right) \frac {v v_S}{\Lambda^2}  \nn \\
\frac{\delta v_{T2}}{v_T}&=&-\left( \frac{t_4}{2h_2}+\frac {2t_3}{h_1} \right) \frac {v v_S}{\Lambda^2} 
+\left( \frac {t_1}{h_1} -\frac{t_2}{h_2}\right) \frac {v u}{\Lambda^2} \nn \\
 \frac{\delta v_{T1}}{v_T}&=&\left( \frac{t_4}{4h_2}+ \frac {t_3}{h_1} \right) \frac {v v_S}{\Lambda^2} \nn
\eea
Observe that the shifts in three components are different 
but all of the same order of magnitude, as claimed in the text:
 $$\frac{\delta v_{Ti}}{v_T} \sim O(\lambda^2_C)~.$$

\end{document}